\documentclass[10pt,conference]{IEEEtran}
\IEEEoverridecommandlockouts
\usepackage{cite}
\usepackage{amsmath,amssymb,amsfonts}
\usepackage{algorithmic}
\usepackage{graphicx}
\usepackage{textcomp}
\usepackage{bmpsize}
\usepackage{xcolor}
\usepackage{lipsum}
\usepackage{booktabs}
\usepackage{multirow} 
\usepackage{csquotes}
\usepackage{framed}
\usepackage{subcaption}
\usepackage[colorinlistoftodos]{todonotes}

\usepackage[colorlinks=true,urlcolor=black]{hyperref}
\def\BibTeX{{\rm B\kern-.05em{\sc i\kern-.025em b}\kern-.08em
    T\kern-.1667em\lower.7ex\hbox{E}\kern-.125emX}}
\usepackage[capitalise,noabbrev]{cleveref}
\begin{document}

\title{A Lightweight Plug-in Module for Introducing Post-Quantum Cryptography in Higher Education: An Experience Report}

\author{
\IEEEauthorblockN{
Ainaz Jamshidi, Khushdeep Kaur, Karen Chen, Aryya Gangopadhyay, Lei Zhang
}
\IEEEauthorblockA{
Department of Information Systems, University of Maryland, Baltimore County\\
Baltimore, MD, USA\\
\{ainazj1, khushdk1, lujiec, gangopad, leizhang\}@umbc.edu
}
}

\maketitle

\begin{abstract}

Post-quantum cryptography (PQC) is becoming an increasingly important topic for computing education, yet integrating PQC into existing curricula is challenging because it draws on programming, mathematics, cryptography, and quantum concepts. In this experience report, we describe the design, implementation, and pilot deployment of a one-week lightweight PQC plug-in module embedded in an undergraduate cybersecurity course and a graduate software engineering course. The module is designed to introduce core PQC concepts, motivate quantum threats to current cryptographic systems, and connect PQC migration to secure software engineering practice without requiring a standalone course.

We report our experience from four pilot deployments: two undergraduate and two graduate course offerings. At each academic level, one offering used faculty-led lectures (FLL), while the other used hybrid lectures (HL) combining student-led seminars with faculty-led lectures. Both formats also incorporated hands-on factorization activities and gamified formative assessment. Using Kahoot quiz scores and anonymous student feedback, we observed generally positive student perceptions. The graduate HL cohort achieved a higher mean Kahoot score than the graduate FLL cohort, providing preliminary evidence that the hybrid format may be helpful for graduate students with heterogeneous backgrounds. However, given the short module duration, small sample sizes, and non-randomized course settings, we interpret these results as exploratory rather than conclusive. We therefore emphasize practical lessons learned, design trade-offs, and reusable instructional materials for educators interested in introducing PQC into existing computing courses. This work contributes an adaptable teaching model and open educational resources for broadening PQC awareness and supporting future curriculum development in quantum-safe cybersecurity and software engineering.

\end{abstract}

\begin{IEEEkeywords}
post-quantum cryptography education, lightweight learning module, active learning, student-led seminar, experience report
\end{IEEEkeywords}

\section{Introduction}

Quantum computing is expected to create new opportunities in areas such as optimization, artificial intelligence, and scientific computing, but it also introduces a significant threat to current cryptographic systems. Shor's algorithm can break widely used public-key cryptographic schemes such as Rivest–Shamir–Adleman (RSA), elliptic-curve cryptography (ECC), and Diffie--Hellman (DH) key exchange~\cite{shor1999polynomial}. Grover's algorithm can also reduce the effective security strength of symmetric-key schemes, requiring larger key sizes for equivalent protection~\cite{grover1996fast}. Although cryptographically relevant quantum computers are not yet available, the cybersecurity community has emphasized the urgency of preparing for ``harvest-then-decrypt'' attacks, in which adversaries collect encrypted data today and decrypt it once sufficiently powerful quantum computers become available~\cite{zhang2020quantum, zhang2023making, mashatan2021complex}.

Post-quantum cryptography (PQC) provides a practical path toward quantum-safe security because it can be deployed on classical computing infrastructure. In August 2024, the National Institute of Standards and Technology (NIST) released the first finalized PQC standards, including ML-KEM for key encapsulation and ML-DSA and SLH-DSA for digital signatures~\cite{BibEntry2024Aug}. At the same time, government and industry initiatives, such as the NSA Commercial National Security Algorithm (CNSA) Suite 2.0, have established timelines for transitioning national security systems to PQC~\cite{CSACNSA210:online}. These developments make PQC migration not only a cryptographic problem but also a software engineering and cybersecurity challenge that requires cryptographic inventory, library replacement, protocol redesign, system testing, and long-term crypto-agility~\cite{zhang2020quantum, zhang2023making, ott2019identifying, wiesmaier2021pqc, nather2024migrating}.

However, the growing policy and technical urgency of PQC has not yet been matched by sufficient workforce preparation. Recent studies suggest that many organizations remain at an early stage of PQC readiness, with limited migration roadmaps and immature tooling~\cite{ahmed2025survey, erikssonquantum, le2025enterprises}. A related challenge appears in computing education: current curricula often emphasize traditional cryptography theory, general cybersecurity concepts, or quantum computing foundations, but provide limited opportunities for students to learn PQC as an applied cybersecurity and software engineering topic~\cite{jamshidi2024let, borrelli2024designing}. This creates a need for educational approaches that can introduce PQC concepts without requiring a full standalone course.

\begin{figure}[thp!]
    \centering
    \includegraphics[width=0.95\linewidth]{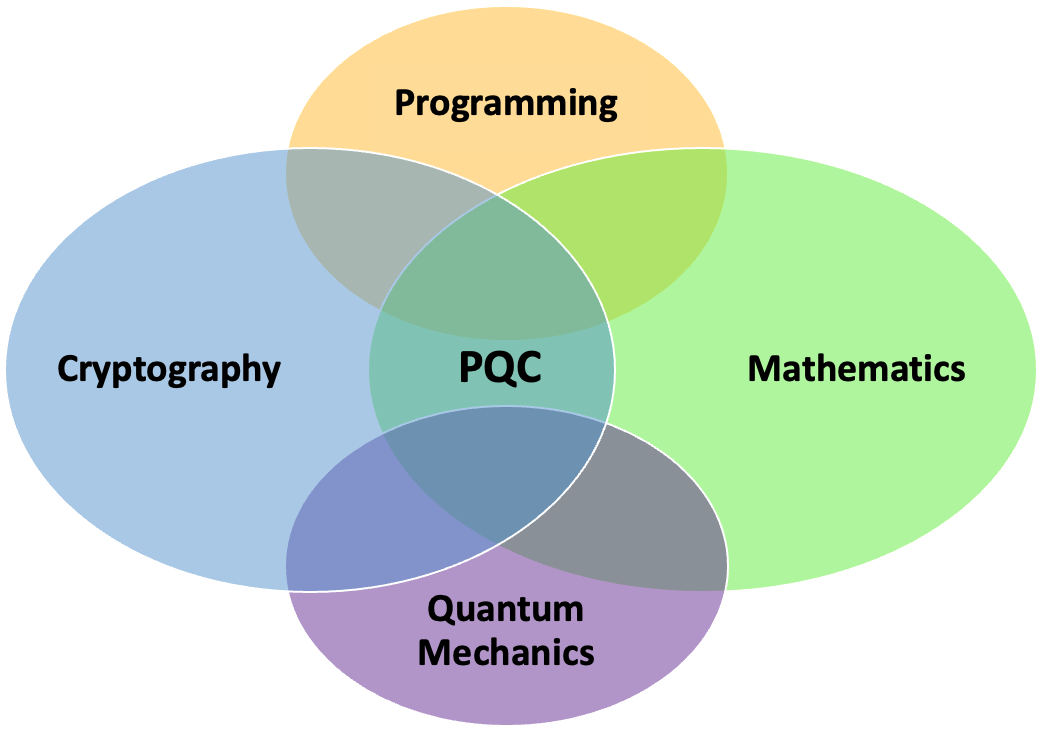}
    \caption{The interdisciplinary nature of PQC competency, which includes: 1) programming skills to implement and apply PQC algorithms; 2) mathematical foundations for understanding key exchange, key encapsulation, digital signatures, and PQC algorithms; 3) cryptography knowledge for understanding encryption protocols and procedures; and 4) quantum mechanics concepts for understanding quantum threats.}
    \label{fig:pqc}
\end{figure}

Teaching PQC is challenging because the topic is inherently interdisciplinary. As shown in Figure~\ref{fig:pqc}, PQC learning draws on programming, mathematics, cryptography, and quantum mechanics. Students may need to understand public-key encryption and digital signatures, basic number theory and lattice-based assumptions, quantum threats such as Shor's and Grover's algorithms, and practical software migration issues. In many computing programs, however, students have uneven preparation across these areas. Moreover, undergraduate and graduate curricula are already crowded, making it difficult to add a complete PQC course even when the topic is recognized as important.

Several educational efforts have begun to address related needs. Some computing programs offer quantum computing courses, such as Georgia Institute of Technology's CS 8803~\cite{GT_CS8803} and the University of Warwick's CS 419~\cite{WarwickCS419}, which provide foundations in quantum algorithms and computation. Other courses, such as Florida Atlantic University's graduate course on the mathematical foundations of PQC~\cite{FAU2026}, provide rigorous theoretical preparation. Professional training opportunities have also emerged, including the International Telecommunication Union (ITU) Academy's 20-hour online training program~\cite{ITU_PQC_2025} and Perpetual Solutions' three-day PQC course~\cite{PerpetualSolutions_PQC_3Day}. While these efforts are valuable, they do not fully address the need for lightweight, reusable modules that can be embedded into existing higher-education computing courses to introduce PQC awareness, core concepts, and migration-oriented thinking.

To address this gap, we design a one-week lightweight PQC plug-in module that can be integrated into existing computing courses. Rather than attempting to provide comprehensive PQC training within a short period, the module aims to raise awareness, introduce foundational concepts, and help students connect PQC to cybersecurity and software engineering practice. The module is piloted in two course contexts: an undergraduate cybersecurity analytics course and a graduate software engineering course. Because of the short duration and the exploratory nature of the deployment, we position this paper as an experience report rather than a controlled evaluation of PQC learning effectiveness.

This paper reports our experience designing, implementing, and iterating the PQC plug-in module. The module combines concise lectures, hands-on activities, student-led seminars, and gamified formative assessment to help students engage with technically dense material in a compressed instructional setting. We use student feedback and Kahoot-based~\cite{KahootLe13:online} formative assessment results as preliminary evidence of student engagement and short-term conceptual understanding, while recognizing the limitations of small cohorts, non-randomized course settings, and a one-week intervention.

The \textbf{contributions} of this paper are as follows:
\begin{enumerate}
    \item We present the design of a lightweight PQC plug-in module that introduces quantum threats, PQC concepts, and PQC migration issues within existing undergraduate and graduate computing courses.
    \item We describe our experience deploying the module across four pilot offerings in cybersecurity and software engineering contexts, highlighting how the module was adapted for students with different backgrounds.
    \item We report preliminary observations from formative assessment and student feedback, and summarize practical lessons learned for educators who wish to introduce PQC into already crowded computing curricula.
    \item We provide open educational materials, including lecture videos, slides, and assessments, to support reuse and adaptation by other instructors. Our artifacts are publicly available at \url{https://doi.org/10.5281/zenodo.13909016}, and our courseware is available at \url{https://est.umbc.edu/teaching/pqc-teaching-module}.
\end{enumerate}

Overall, this work contributes a practical and reusable model for introducing PQC into higher education. Its primary \textbf{goal} is not to claim that a one-week module can produce deep PQC expertise, but to show how a short, carefully designed instructional unit can help students begin developing quantum-safe security awareness and prepare for deeper study of PQC and secure software migration.

The remainder of this paper is organized as follows. Section~\ref{sec:method} presents the design of the lightweight PQC plug-in module, including its integration into two existing computing courses, the core module content, and the active-learning components. Section~\ref{sec:PilotMethod} describes the pilot deployments, student cohorts, and data collection methods used to document our experience. Section~\ref{sec:results} reports preliminary observations from Kahoot-based formative assessments and student feedback. Section~\ref{sec:takeaways} discusses lessons learned from the pilots and reflects on the limitations of the study. Finally, Section~\ref{sec:conclusions} concludes the paper and outlines directions for future curriculum development and evaluation.

\section{Module Design}\label{sec:method}

This section describes the design of the one-week lightweight PQC plug-in module. Because the module is intended to fit into existing computing courses rather than function as a standalone PQC course, our design focuses on three goals: 1) identifying suitable insertion points in existing curricula, 2) selecting a compact set of foundational PQC concepts that could be taught within a short instructional window, and 3) incorporating active-learning activities to support engagement with interdisciplinary and technically dense material. We first describe the module integration strategy in Section~\ref{module_strategy}. We then present the core module structure in Section~\ref{subsec:lecture}, followed by the active-learning strategies used in the pilots in Section~\ref{subsec:active}.

\subsection{Module Integration Strategy}\label{module_strategy}

\begin{figure}[thp!]
    \centering
    \includegraphics[width=0.95\linewidth]{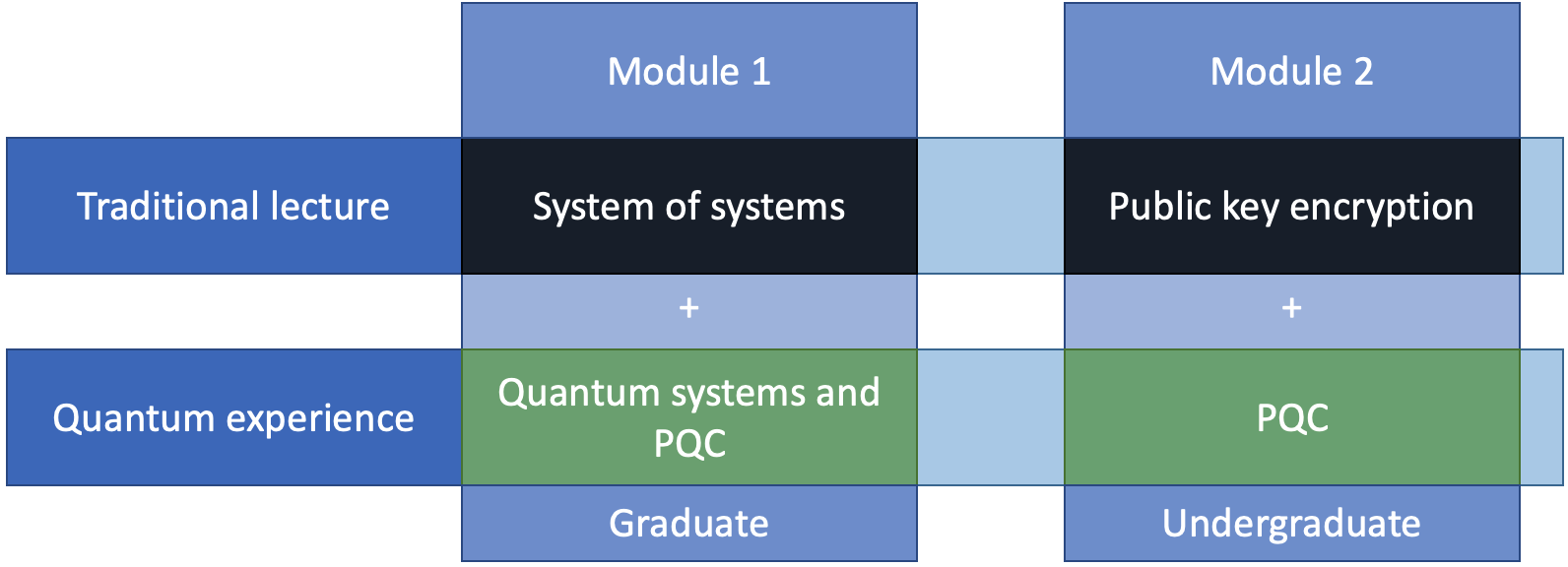}
    \caption{Integration of PQC learning modules into existing curricula at undergraduate and graduate levels.}
    \label{fig:lectures}
\end{figure}

The main design goal of the plug-in module is to introduce PQC without requiring a new standalone course or major changes to an existing syllabus. We therefore embed the module into two courses at the Department of Information Systems (IS) that already contained natural entry points for PQC: a graduate-level systems analysis and design course (IS 636) and an undergraduate-level cybersecurity analytics course (IS 471). These two contexts allow us to examine how the same core PQC content could be adapted for students with different backgrounds and learning goals. Figure~\ref{fig:lectures} illustrates the high-level placement of the module within the two curricula.

\textbf{IS 636: Structured Systems Analysis and Design (graduate).}
IS 636 is a core graduate course that introduces students to the analysis and design of information systems from a systems-development life cycle (SDLC) perspective. The course covers the role of the systems analyst, major SDLC phases, and techniques for requirements analysis, system modeling, and design. 

The semester is organized around topics that closely follow a standard systems analysis and design textbook and a team-based project. Early weeks introduce the SDLC, systems analysts’ responsibilities, project initiation, and feasibility analysis. Subsequent weeks focus on project planning and management, followed by a multi-week analysis phase covering requirements engineering, functional and non-functional requirements, requirements elicitation, and analysis. The design portion of the course includes use case modeling and UML, process modeling, data flow diagrams, data modeling, architecture design, and user interface, hardware, and software design.

The PQC module is integrated late in the semester, during the ``System of Systems'' topic. At this point, students learn complex information systems and inter-system interactions, computer architectures, cloud computing, and design trade-offs for distributed architectures. Building on students' prior exposure to SDLC, architecture design, and system-of-systems concepts, the PQC session frames PQC as a cross-cutting concern that affects requirements, architecture, and implementation decisions in large-scale enterprise environments.

\textbf{IS 471: Data Analytics for Cybersecurity (undergraduate).} 
IS 471 is an upper-level undergraduate course that provides a foundations-oriented introduction to cryptography, real-world security protocols, and malware analysis, with a strong emphasis on hands-on work using open-source tools. Throughout the semester, students implement and experiment with cryptographic primitives and security analytics.

The course is organized into fifteen instructional modules that alternate between lecture and lab. After an initial orientation, students work through a sequence of classical cryptographic topics, i.e., basic cryptographic algorithms, substitution ciphers (with a dedicated lab), stream ciphers (two weeks with labs), block ciphers (lectures and labs), and a midterm exam. The second half of the course introduces asymmetric cryptography in two lecture–lab pairs, cryptographic hashing and corresponding labs, and real-world security protocols (e.g., TLS-style scenarios) with practical lab exercises. The course concludes with ``Advanced topics in cybersecurity'' and a final exam.

In this context, the PQC module is positioned as the ``Advanced Topics'' after students have completed the two-week unit on asymmetric cryptography and have seen how classical public-key primitives are used in real-world security protocols. Building on their experience implementing RSA-like schemes and working with key management and hashing, the PQC lecture revisits those protocols through the lens of quantum threats and post-quantum replacements. This placement allows the module to leverage students’ freshly acquired understanding of public-key encryption, digital signatures, and protocol design, while introducing PQC as an extension of, rather than a replacement for, the cryptographic concepts already covered in the course.

Together, IS 636 and IS 471 provide complementary curricular contexts for piloting the PQC plug-in module. The graduate course foregrounds systems analysis, architecture, and system-of-systems reasoning, while the undergraduate course foregrounds cryptography, protocols, and hands-on security analytics. This pairing allows us to explore how a shared lightweight PQC module could be adapted across course levels and student backgrounds while preserving a common core of PQC concepts and active-learning activities.

\subsection{Module Structure}\label{subsec:lecture}

\begin{table*}[thp!]
    \centering
    \caption{Comparison of undergraduate and graduate courses based on cybersecurity and structured software systems topics.}
    \label{tab:course_comparison}
    \begin{tabular}{@{}llp{6cm}p{6cm}@{}}
    \toprule
    \textbf{} & \textbf{Undergraduate (IS 471)} & \textbf{Graduate (IS 636)} \\
    \midrule
    \textbf{Course title} & Data Analytics for Cybersecurity & Structured Systems Analysis and
Design \\
    \textbf{Prerequisite} & All learned public/private-key encryption & May or may not have taken cybersecurity courses before \\
    \textbf{PQC-related topics} & PQC, Cryptocurrency & System of Systems, Cloud-based quantum computing, PQC \\
    \textbf{Lecture length} & 1 hour and 15 minutes & 2 hours and 30 minutes \\
     \textbf{Preparation time} & 1--2 hours & 1--2 hours \\
    \textbf{Student cohort size} & 24 [12 (Spring 2023) and 12 (Spring 2024)] & 28 [15 (Spring 2023) and 13 (Fall 2023)] \\
    \textbf{Cohort background} & Seniors in Information Systems & Diverse students (mostly international) from STEM and non-STEM disciplines \\
    \bottomrule
    \end{tabular}
\end{table*}

The one-week PQC module is organized into four sessions, where we 1) review classical cryptography and motivate the quantum threat; 2) introduce essential quantum computing concepts (such as superposition and collapse);  3) present Shor's algorithm, PQC algorithms; and 4) wrap up with a concrete PQC migration case study. The graduate module allocates more time to classical cryptography, quantum computing fundamentals, and system-of-systems implications, whereas the undergraduate module compresses these introductions and emphasizes concrete PQC schemes and migration guidelines, given the difference in background knowledge between students enrolled in those two courses. The differences between the graduate and the undergraduate courses can be seen in Table~\ref{tab:course_comparison}.

\textbf{Session 1. From classical cryptography to quantum threats.} 
The module starts with a compact review of classical cryptography to reactivate students’ prior knowledge and create a shared baseline. We contrast symmetric and asymmetric primitives (e.g., AES, 3DES, RSA, ECC, Diffie–Hellman), discuss security strength and key-size recommendations, and connect these mechanisms to concrete protocols such as TLS handshakes. 

\textbf{Session 2. Quantum-computing primer.} 
The second session provides a targeted quantum-computing primer tailored to security and systems students with heterogeneous backgrounds. We introduce qubits, superposition, entanglement, and measurement using simple visual examples, and contrast them with classical bit-level computation. A short historical perspective on quantum hardware and the complexity class of bounded-error quantum polynomial time, which helps students understand why quantum algorithms challenge existing security assumptions. We then give an intuitive walkthrough of Shor’s factoring algorithm and Grover’s search algorithm, focusing on the structure of the attack rather than mathematical formalism. The discussion explicitly links these results to real-world risks such as ``harvest-then-decrypt'' attacks and long-lived secrets (e.g., national security data). In the graduate course, this is complemented by a brief introduction to Dirac notation and cloud-based quantum platforms (e.g., IBM-style circuit execution~\cite{wille2019ibm}), whereas the undergraduate version relies on high-level circuit diagrams and analogies without assuming linear-algebra fluency.

\textbf{Session 3. PQC schemes.} 
After establishing the threat model, the lecture shifts to defensive measures. We distinguish quantum cryptography (e.g., quantum key distribution) from PQC (classical algorithms designed to resist quantum adversaries) and motivate why current standards bodies and national strategies prioritize PQC for near-term deployment. We introduce the main NIST-standardized schemes at a high level, such as CRYSTALS-Kyber for key encapsulation and CRYSTALS-Dilithium and SPHINCS+ for digital signatures~\cite{BibEntry2024Aug}, and show how they can replace or augment RSA, ECC, and DH in existing protocol stacks. Policy milestones (e.g., NIST PQC standardization, NSA CNSA 2.0, U.S. federal PQC migration memorandum) are used to frame PQC as an immediate security and compliance concern rather than a purely theoretical topic.

\textbf{Session 4. A PQC migration case study.} 
To connect these ideas to real software, we use an industrial migration case study (IBM Db2 database management software~\cite{zhang2023making}) and walk students through the end-to-end authentication and TLS stack. Students identify where DH and RSA appear, then reason about how ML-KEM key encapsulation and PQC signatures could be integrated into an existing cryptographic library and deployment pipeline. We structure this discussion using a ``7E-style'' roadmap (e.g., Engage, Examine, Evolve, Educate, Estimate, Execute, Evaluate)~\cite{zhang2023making}, emphasizing that PQC migration is a multi-stage software-evolution problem impacting requirements, architecture, performance, and interoperability.

At the end of the module, a short quiz (administered via Kahoot~\cite{KahootLe13:online}) checks basic understanding in 1) the distinction between symmetric and asymmetric cryptography, 2) quantum mechanics, 3) the high-level impact of Shor’s algorithms, 4) the definition of PQC, and 5) quantum superposition. We use the same quiz across iterations, allowing us to compare learning outcomes across cohorts and various pedagogical strategies.

\subsection{Active Learning Strategy}\label{subsec:active}

Active learning, as a pedagogical strategy, has been widely recognized as an effective teaching approach in STEM education~\cite{cundell2018student, van2023remote, aji2019impact}. In computer science, active learning methods such as peer-led team learning~\cite{biggers2009using, alo2007work, caceffo2018exploring} and project-based learning~\cite{stewart2010using} have been shown to improve student engagement, problem-solving skills, and conceptual retention. The effectiveness of student-led discussions~\cite{minhas2012effects, kurczek2014student, casteel2007goodbye, kassab2005student} has also been demonstrated, particularly in interdisciplinary fields, where students benefit from collaborative knowledge sharing.

With recent breakthroughs in quantum computing, educational research has begun to address quantum-related topics \cite{salehi2021computer, fox2020preparing, nita2023challenge}. However, a notable gap remains regarding the integration of active learning approaches into quantum cryptography education, particularly for PQC. The interdisciplinary nature of quantum cryptography---requiring foundational knowledge in linear algebra, cybersecurity, and quantum mechanics---poses significant challenges. Adding to this complexity is the limited time available to introduce PQC concepts effectively and efficiently within already crowded computing curricula. These challenges underscore the need for pedagogical approaches that both engage students and manage cognitive load; consequently, active learning offers a particularly promising strategy for supporting meaningful PQC learning within a compact instructional module by moving beyond a lecture-centric delivery format. 

We explore several active-learning components in the PQC module as follows.

\textbf{Student-led seminars.} 
The student-led seminars are inspired by the flipped classroom approaches, which reverse the traditional teaching model by requiring students to engage with instructional materials before class (e.g., video lectures, readings) and participate in active discussions and problem-solving sessions during class~\cite{chen2017using, freeman2014active}. This model has been successfully applied in software engineering education~\cite{lin2021effects}, improving student engagement and independent learning skills. However, one drawback is that students with weaker foundational knowledge may struggle to grasp complex material before classroom discussions, leading to disengagement. 
In those sessions, the lecture begins with short student-led seminars. Students (either as a group or individually) can pick any quantum-computing-related topics (e.g., ``Shor’s algorithm,'' ``Superposition,'' ``PQC,'' and ``Quantum hardware'') one week in advance. Students then prepare an ``elevator-pitch'' style presentation (typically 1--2 minutes, without any slides) and deliver it at the beginning of the next class.
To avoid overloading students, we provide a one-page guideline including a small set of topics and questions to address, e.g., ``What is a qubit?”. After the seminar, we reserve time for a short ``Q\&A'' and a brief instructor-led synthesis, where we correct misconceptions and connect the topic to the rest of the module. 

\begin{figure}[thp!]
    \centering
    \includegraphics[width=0.95\linewidth]{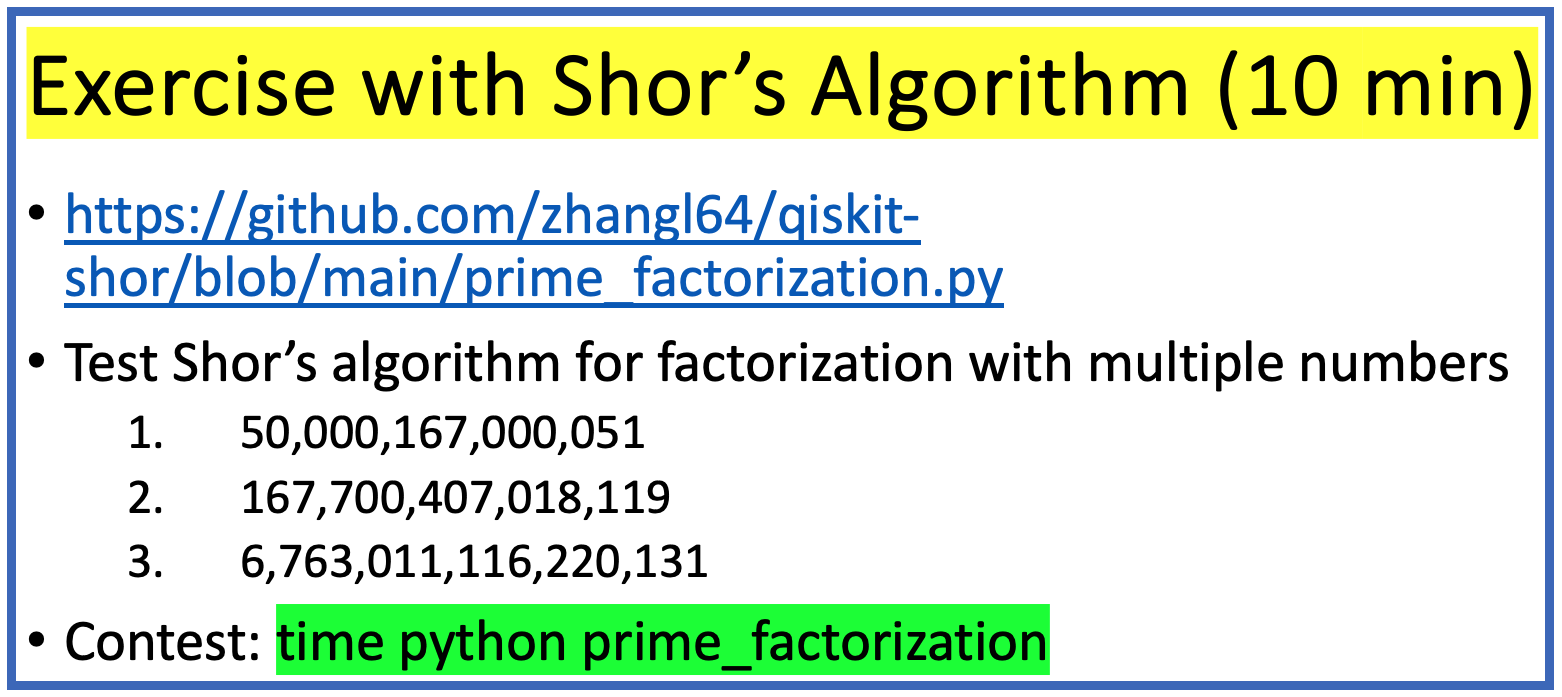}
    \caption{An example of a hands-on exercise in the slides. Participants test their implementations of integer factorization with numbers in various orders of magnitude (those who successfully complete the tasks more quickly are awarded higher scores).}
    \label{fig:coding}
\end{figure}

\textbf{Hands-on practice: integer factorization and the complexity of RSA.} A second pillar of our active learning design is a 20-minute hands-on exercise on integer factorization (as shown in Figure~\ref{fig:coding}). The goal is not to teach number theory but to make the computational hardness assumptions behind RSA concrete. Students receive a GitHub repository with starter code that has implemented naive trial division for factoring integers of increasing size (Figure~\ref{fig:coding}). Working individually, students are asked to

\begin{enumerate}
\item run the factorization on a sequence of inputs and record the measured running time for each iteration;
\item compare the running time with each other for a fun competition; 
\item reflect on why realistic RSA key sizes (e.g., 2048 bits, which is much larger than the integer in our exercise) justify current security claims in a classical setting.
\end{enumerate}

In the graduate course, students are additionally asked to relate these observations to Big-$O$ notation and to discuss how quantum computing has added bounded-error quantum polynomial time (BQP) to the complexity zoo. In both settings, the instructor used the exercise debrief to revisit the ``harvest-then-decrypt'' threat and to motivate PQC as a defense against future adversaries.

\textbf{Gamified Formative Assessment.} Gamification has gained traction in STEM education as an effective way to enhance motivation and engagement~\cite{ivanova2019gamification}. Techniques such as game-based coding challenges, leaderboards, and interactive quizzes~\cite{videnovik2023game} have been used to make learning more immersive and competitive. In PQC education, we incorporate Kahoot-based assessments to introduce an element of gamification, providing students with instant feedback and an interactive learning experience. However, while gamified learning focuses on engagement and retention, it does not inherently foster deeper conceptual discussions. By integrating student-led seminars, we enable students to actively debate, analyze, and contextualize PQC principles rather than passively absorbing gamified content.

\begin{figure}[thp!]
    \centering
    \includegraphics[width=0.95\linewidth]{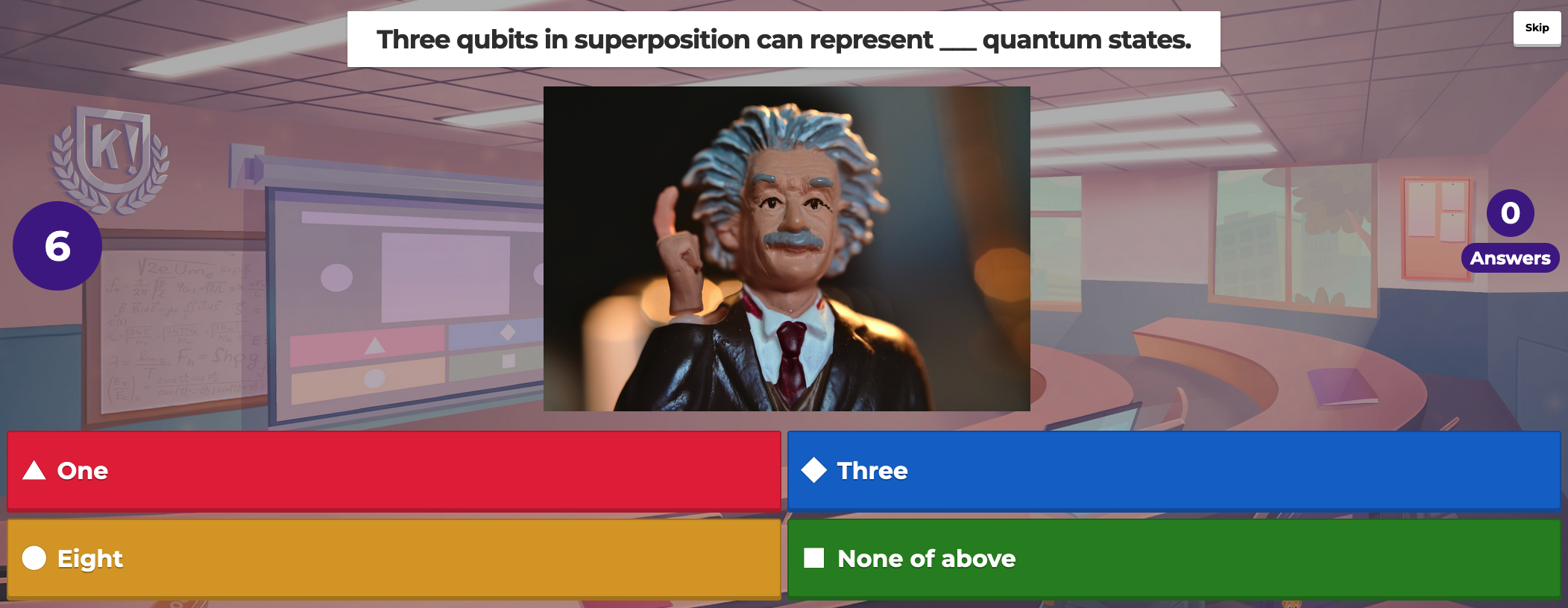}
    \caption{A multiple-choice question in Kahoot. The question is projected on the big screen in the classroom, and participants will select the correct answer on their mobile devices. Participants who can choose the correct answer faster receive higher scores.}
    \label{fig:kahoot}
\end{figure}

We use gamified quizzes delivered via Kahoot as a formative assessment implemented in all sessions. Each PQC session concludes with a five-question Kahoot game to check  understanding of PQC-related concepts (e.g., distinguishing symmetric and asymmetric cryptography, see Figure~\ref{fig:kahoot}). Students answer the questions individually on their own devices, with immediate feedback and a live leaderboard providing a light competitive element. After each question, the instructor briefly explains the correct answer, highlights common distractors, and explicitly ties the concept back to the lecture and hands-on activities. Because the same core question set is used across cohorts and formats, these quizzes served a dual purpose: 1) they function as retrieval practice and engagement tools in the classroom, and 2) they provide a standardized measure for comparing students’ conceptual understanding of PQC across contexts.

\section{Pilot Studies}\label{sec:PilotMethod}

This section describes the four pilot deployments of the PQC plug-in module. Consistent with the experience-report framing of this paper, our goal is to document how the module has been implemented in real course settings, how the instructional format has evolved across iterations, and what forms of formative assessment and student feedback have been collected. We first describe the course contexts and student cohorts, and then summarize the data collection and analysis procedures used to support our preliminary observations.

\subsection{Course Contexts and Student Cohorts}

\begin{table}[thp!]
    \centering
    \caption{Structure of learning modules with active learning activities. We introduce active learning activities, such as hands-on coding exercises and Kahoot in the first iteration, and we add student-led seminars (i.e., Hybrid Lecture) as an extra active learning approach in the second iteration. FLL: Faculty-Led lecture; HL: Hybrid Lecture}
    \label{tab:structure}
    \begin{tabular}{@{}lll@{}}    
    \toprule
    \textbf{Iteration} & \textbf{Graduate (IS 636)} & \textbf{Undergraduate (IS 471)} \\
    \midrule
    \multirow{1}*{1} & Spring 2023: FLL & Spring 2023: FLL\\ 
    \midrule
    \multirow{1}*{2} & Fall 2023: HL & Spring 2024: HL \\ %
    \bottomrule
    \end{tabular}
\end{table}

As shown in Tables~\ref{tab:course_comparison} and \ref{tab:structure}, we conduct two iterations of implementation comprising a total of four pilots: two in the undergraduate cybersecurity course (IS 471)  and two in the graduate software engineering course (IS 636). The graduate lecture is two and a half hours, and the undergraduate lecture is one hour and fifteen minutes. In the graduate lecture, we have more introductory materials on quantum systems because of the length of the graduate lecture. In contrast, the undergraduate lecture mainly focuses on PQC. The two iterations employ slightly different active learning strategies. While both include hands-on coding activities and Kahoot quizzes, the second iteration adds a student-led seminar as a flipped-classroom implementation.

We conduct a comparative study focusing on two instructional formats, i.e., faculty-led lectures (FLL) and hybrid lectures (HL, a combination of student-led seminars and faculty-led lectures), respectively. We delivered PQC lectures over two iterations between Spring 2023 and Spring 2024 (see Table~\ref{tab:structure}). The first iteration used FLL, and the second iteration used HL. Both iterations contain multiple active learning approaches, such as hands-on coding exercises and Kahoot games. However, only the second iteration introduces student-led seminars prior to faculty-led lectures. We assess the impact of both formats on students’ learning and engagement through the results of Kahoot quizzes and students' subjective feedback via anonymous surveys.

\textbf{First Iteration:} In the spring semester of 2023, we developed our first PQC learning modules with active learning approaches, such as hands-on coding exercises and Kahoot quizzes, for IS 636 and IS 471 (see Table~\ref{tab:structure}). An example of hands-on exercises is shown in Figure~\ref{fig:coding}, and an example of Kahoot multiple-choice questions can be seen in Figure~\ref{fig:kahoot}. 

\textbf{Second Iteration:} After completing the initial PQC learning modules, we observed several learning challenges in the graduate course (i.e., low Kahoot scores). To increase student learning outcomes, we integrated a \textbf{student-led seminar} into the PQC learning module and created the new hybrid learning PQC module. We implemented the second iteration of the PQC learning modules in the same courses in the fall semester of 2023 and the spring semester of 2024. Both the graduate and undergraduate courses follow the same structure. The new PQC lecture begins with a student-led seminar, followed by the faculty-led lecture with hands-on coding exercises, and concludes with a Kahoot quiz  (see Table~\ref{tab:structure}). For student-led seminars, participants choose a quantum computing topic (e.g., ``What is a qubit?'') from a provided list or suggest their own and deliver an elevator pitch individually or in groups.

\subsection{Data Collection and Analysis}\label{subsec:data}

\paragraph{\textbf{Formative Assessment}}

To provide a consistent measure of short-term conceptual understanding, the same set of Kahoot questions are used across the PQC lectures. The quiz includes four true-or-false questions and one multiple-choice question covering key concepts introduced in the module. Kahoot is used primarily as a formative assessment and engagement tool, providing immediate feedback to students and instructors. We report descriptive statistics for Kahoot scores and use a $t$-test~\cite{student1908probable} to explore differences between FLL and HL cohorts. Because the cohorts are small and non-randomized, the statistical results are interpreted as preliminary indicators rather than definitive evidence of learning effectiveness. More detailed results and analysis can be found in Section~\ref{subsec:kahoot}.

\begin{table*}[thp!]
    \centering
    \caption{Student lecture evaluation.}
    \label{tab:feedback}
    \resizebox{\textwidth}{!}{\begin{tabular}{@{}lll@{}}    
    \toprule
    \textbf{Question \#} & \textbf{Question} & \textbf{Choice} \\
    \midrule
    1 & Lecture pace is: & 1. Too slow 2. Just right 3. Too fast  \\ 
    2 & Amount of material presented is: & 1. Too little 2. Just right 3. Too much \\
    3 & Difficulty of material presented is: & 1. Too easy 2. Just right 3. Too hard  \\
    4 & Is the topic interesting? & 1. Agree 2. Neutral 3. Disagree  \\
    5 & Does the instructor explain the material clearly? & 1. Strongly agree 2. Agree 3. Neutral 4. Disagree 5. Strongly disagree  \\
    6 & What do you like about the lecture? & Open-ended question  \\
    7 & What do you dislike about the lecture? & Open-ended question  \\
    8 & Your general comments about the lecture: & Open-ended question  \\
    \bottomrule
    \end{tabular}}
\end{table*}

\paragraph{\textbf{Student Evaluation}}

To capture students' perceptions of the module, we administer anonymous end-of-lecture surveys using Google Forms. The survey asks students to evaluate lecture pace, amount of material, difficulty, topic interest, and clarity, and also includes open-ended questions about what students liked, disliked, and suggested for improvement. Table~\ref{tab:feedback} summarizes the survey questions. To compare the distributions of closed-ended survey responses between the FLL and HL cohorts within each academic level, we use Pearson’s chi-square test~\cite{mchugh2013chi} of independence. When a comparison has one degree of freedom, Yates's continuity correction~\cite{yates1934contingency} is applied; otherwise, no continuity correction is used. The test is not performed when all participants select the same response category because the resulting table has zero degrees of freedom. Statistical significance is evaluated at $\alpha = 0.05$. Given the small cohort sizes and low frequencies in some response categories, the results are interpreted as exploratory. Results and analysis can be found in Section~\ref{subsec:evaluation}.

\section{Results}\label{sec:results}

This section reports preliminary observations from the four pilot deployments. We summarize Kahoot-based formative assessment results and student feedback for the two instructional formats: FLL and HL. 

\subsection{Student Learning Outcomes}\label{subsec:kahoot}

In the undergraduate course, both the FLL and HL groups have 12 students. For the graduate course, 15 students are enrolled in the FLL group, while 13 students participate in the HL group.  All the undergraduate students take Kahoot quizzes. However, one student in the graduate HL group misses the quiz. The results of the Kahoot quizzes for each group are reported as the mean $ \pm $ standard deviation (SD), along with the coefficient of variation (CV), which provides a measure of the relative variability in students' performance. The statistical significance of the differences between groups is assessed using the $p$-value, with values below 0.05 considered statistically significant. 

\begin{table}[thp!]
    \centering
    \caption{The evaluation of students' performance in Kahoot quizzes under the practice of FLL and HL groups (the higher, the better). The student scores are reported as mean $ \pm $ standard deviation. The $p$-value for the $t$-test is used to determine the statistical significance of the differences between the studied groups, and the CV shows the relative variability of the scores.}
    \label{tab:performance}
    \begin{tabular}{@{}llrrr@{}}
    \toprule
    \textbf{Course} & \textbf{Group} & \textbf{Score Mean $ \pm $ SD} & \textbf{CV} & \textbf{\textbf{$P$}-value} \\
    \midrule
    \multirow{2}*{Undergraduate} & FLL & 3540.84 $ \pm $ 1006.96 & 0.28 & \multirow{2}*{0.367} \\
        & HL & 5216.84 $ \pm $ 5954.70 & 1.14 &  \\
    \midrule
    \multirow{2}*{Graduate} & FLL & 2759.20 $ \pm $ 1159.96 & 0.42 & \multirow{2}*{0.046*} \\
        & HL & 3683.91 $ \pm $ 1009.41 & 0.27 &  \\
    \bottomrule
    \multicolumn{5}{l}{\footnotesize{*\( p < 0.05 \) is considered statistically significant.}} \\
    \end{tabular}
\end{table}

\textbf{Kahoot Scores}. Kahoot scores are awarded based on the speed of the correct answers~\cite{BibEntry2025Jan}. Single-select questions offer up to 1,000 points when a player responds correctly. As presented in Table~\ref{tab:performance}, the HL group in the undergraduate course achieves a higher mean score (5216.84 $ \pm $ 5954.70) compared to the FLL group (3540.84 $ \pm $ 1006.96), though this difference is not considered statistically significant ($p$-value = 0.367). In contrast, for the graduate course, the HL group not only has a higher mean score (3683.91 $ \pm $ 1009.41) but also a lower SD and CV compared to the FLL group (2759.2 $ \pm $ 1159.96), with a $p$-value of 0.046, indicating a statistically significant between-cohort difference favoring the HL group. 
These results provide preliminary evidence that the student-led seminar component may support graduate students’ short-term conceptual understanding, particularly in cohorts with heterogeneous backgrounds. However, because the comparison involved different course offerings rather than randomized groups, the result should be interpreted cautiously.

This improvement is further supported by the increase in Kahoot correctness rates, which have risen from 58\% (FLL) to 78\% (HL). A potential explanation for the low correctness rate in the FLL group is the diverse educational backgrounds of our graduate students (e.g., some of them may graduate from non-STEM disciplines). In contrast, graduate students in the HL group, because of the extra self-directed learning time (to prepare the seminars) before lectures, are better equipped for independent learning than students in the FLL group, as reflected in their higher average Kahoot scores.

In summary, the Kahoot results provide preliminary evidence that the HL approach may support students' short-term conceptual understanding in the graduate course, where the HL cohort achieved higher scores than the FLL cohort with a statistically significant difference. In the undergraduate course, the HL cohort also achieved a higher average score, but the difference was not statistically significant. These results should be interpreted cautiously given the small cohort sizes, the short duration of the module, and the use of Kahoot as a formative assessment tool.

\subsection{Student Feedback}\label{subsec:evaluation}
 
We adopt both statistical and qualitative thematic analysis to analyze the responses collected from the end-lecture questionnaires. The questionnaires consist of five multiple-choice items (lecture pace, amount of material, difficulty, topic interest, and clarity) and three open-ended questions (see Table~\ref{tab:feedback}). We begin by examining the overall statistical comparison between the FLL and HL groups, followed by descriptive analysis of the response distributions and qualitative insights from open-ended feedback.

\begin{table*}[t]
\centering
\caption{Comparison of lecture feedback between FLL and HL formats within undergraduate and graduate cohorts (higher percentages indicate more favorable perceptions). All the response rates for HL groups are 100\%, while the undergraduate and graduate FLL groups have response rates of 92\% and 87\%, respectively.}
\label{tab:feedback_comparison}
\begin{tabular}{l|ccc|ccc}
\toprule
& \multicolumn{3}{c|}{\textbf{Undergraduate}} & \multicolumn{3}{c}{\textbf{Graduate}} \\
\textbf{Metric} & \textbf{FLL} & \textbf{HL} & \textbf{$p$-val} & \textbf{FLL} & \textbf{HL} & \textbf{$p$-val} \\
\midrule
\multicolumn{7}{l}{\textit{\% Just Right}} \\
Lecture pace    & 100\% (11/11) & 83.34\% (10/12)& 0.499 & 69.23\% (9/13)& 84.61\% (11/13) & 0.642 \\
Material amount & 100\% (11/11) & 100\% (12/12)& 1.0 & 61.54\% (8/13)& 84.61\% (11/13) & 0.321 \\
Difficulty      & 100\% (11/11) & 91.67\% (11/12) & 0.965 & 69.23\% (9/13) & 92.31\% (12/13) & 0.320 \\
\midrule
\multicolumn{7}{l}{\textit{\% Agree / Strongly Agree}} \\
Interesting     & 100\% (11/11) & 100\% (12/12) & 1.0 & 61.54\% (8/13) & 92.31\% (12/13) & 0.163 \\
Clarity         & 72.73\% (8/11)& 100\% (12/12)& 0.089 & 92.31\% (12/13) & 92.31\% (12/13)& 0.710 \\
\bottomrule
\end{tabular}

\smallskip
\textit{Note.} Statistical significance assessed using $\chi^2$ tests; threshold $\alpha = .05$.
\end{table*}

\paragraph{\textbf{Analysis of Responses to Closed-Ended Questions}}
Table~\ref{tab:feedback_comparison} provides an aggregate comparison of how students evaluated the PQC lectures under the FLL and HL formats. Across both undergraduate and graduate cohorts, the table summarizes the proportion of students who selected the most favorable response for each metric---lecture pace, amount of material, difficulty, and topic interest. According to Table~\ref{tab:feedback_comparison}, student perceptions are strongly positive across both instructional approaches, with HL consistently exhibiting higher descriptive ratings, particularly in the graduate cohort. Graduate HL students report more favorable impressions of the lecture pace, the amount and difficulty of content, and the overall interest of the topic, reflecting a more engaged learning experience. Undergraduate evaluations also trend positively, with both formats performing similarly well. However, none of the observed differences reach statistical significance. This indicates that while HL yields qualitatively better descriptive feedback, especially among graduate students, the two formats are statistically comparable in terms of students’ perceived learning experience. 

As summarized in Table~\ref{tab:feedback_comparison}, students in both undergraduate and graduate cohorts generally reported favorable perceptions of lecture pace, material amount, difficulty, topic interest, and clarity, with HL often showing descriptively stronger ratings. For lecture pace, 100\% of undergraduate FLL students and 83.34\% of undergraduate HL students rated the pace as “just right,” while 16.67\% of undergraduate HL students felt it was “too fast.” At the graduate level, 69.23\% of FLL students and 84.61\% of HL students rated the pace as “just right,” whereas 30.77\% of FLL students and 15.38\% of HL students considered it “too fast.” For the amount of material, both undergraduate groups unanimously rated it as “just right” (100\%). In the graduate course, 61.54\% of FLL students and 84.61\% of HL students reported the amount as “just right,” while 30.77\% of FLL students and 7.69\% of HL students felt there was “too much” material; 7.69\% in both groups felt there was “too little.” Regarding difficulty, 100\% of undergraduate FLL students and 91.67\% of undergraduate HL students found the material “just right,” with only 8.33\% of HL students rating it as “too hard.” Among graduate students, 69.23\% of FLL students and 92.31\% of HL students rated the difficulty as “just right,” compared with 30.77\% and 7.69\%, respectively, who found it “too hard.” For topic interest, both undergraduate groups reported 100\% agreement that the topic was interesting, while in the graduate course 61.54\% of FLL students and 92.31\% of HL students agreed, with 38.46\% of FLL students and 7.69\% of HL students responding neutrally. In terms of clarity, 100\% of undergraduate HL students either agreed or strongly agreed that the material was clear, compared with 72.73\% of undergraduate FLL students. At the graduate level, 92.31\% of both FLL and HL students either agreed or strongly agreed that the material was clear, while 7.69\% in each group responded neutrally. Overall, these descriptive results suggest that both formats were well received, with HL generally producing more favorable perceptions, particularly in the graduate course.

In summary, the closed-ended survey responses suggest that both FLL and HL formats were generally well received. HL showed descriptively more favorable ratings in several graduate-course measures, including lecture pace, material amount, difficulty, and topic interest. However, these differences were not statistically significant, and therefore should be interpreted as indicative trends rather than conclusive evidence of differences in perceived learning experience (see Table~\ref{tab:feedback_comparison}).

\paragraph{\textbf{Analysis of Responses to Open-Ended Questions}}

For the three open-ended questions (see Table~\ref{tab:feedback}), we conduct a thematic analysis of the students’ responses on what they like or dislike about our lectures and their overall comments or suggestions for improving the lecture. 

For the question ``What do you like about the lecture?'', students' comments highlight two topics in general---1) the lecture presentation/structure and 2) the active learning activities. We observe 72.73\% of undergraduate FLL students indicate that they like the way the lecture is presented, compared to 91.67\% of undergraduate HL students, demonstrating a noticeable increase in satisfaction with the HL format. A similar trend is observed at the graduate level, where 38.46\% of FLL students express satisfaction with the presentation, compared to 53.85\% of HL students. Although satisfaction rates are generally lower among graduate students, HL participants consistently reported higher satisfaction than FLL students at both academic levels.

In the undergraduate course, 45.45\% of FLL students provide positive feedback regarding the active learning method used, while slightly more HL students (50.00\%) report the same. This indicates a fairly similar level of approval for active learning activities between the two groups. In graduate lectures, 23.08\% of HL students provide favorable comments on student-led seminars, and no responses about active learning are collected from the FLL group. This suggests a noticeable difference in how graduate students from the two groups perceive the effectiveness of student-led activities. 

For the question ``What do you dislike about the lecture?'', we do not observe any negative feedback from the participants regarding the presentation style or the active learning methods adopted. Among undergraduate students, 63.64\% of FLL and 58.34\% of HL participants indicate they do not dislike anything about the lecture. Similarly, at the graduate level, 15.38\% of FLL students and 38.46\% of HL students report that they do not dislike any aspect of our lectures in terms of presentation and active learning approach. The remaining comments primarily concern the lecture pace and the amount of material presented.

For the question ``Your general comments about the lecture:'', we manually review the feedback from participants and evaluate their overall learning experience. In the undergraduate group, where attendance rates are 92\% for the FLL group and 100\% for the HL group, both sessions receive universally (100\%) positive feedback regarding the active learning approaches employed or the structure of the HL.  

For example, one undergraduate student noted, 
\begin{displayquote}
``The student presentations at the beginning were an interesting inclusion, as it allowed them to gain a basic understanding beforehand. The Kahoot was also very fun.''    
\end{displayquote} 
Another undergraduate student remarked, 
\begin{displayquote}
``I liked the lecture and it was engaging. I knew very little coming in and I came out learning interesting things.''    
\end{displayquote}

This positive trend extends to the graduate courses as well. In the graduate FLL and HL lectures, with response rates of 60\% and 100\%, respectively, our lectures also receive 100\% positive feedback on the structure of the HL sessions and the active learning strategies. For example, a graduate student emphasized the advantages of our teaching approach as follows.
\begin{displayquote}
``Live demo with the code execution really showed me the impact of quantum computers when our classical computers weren't able to match up to the speed and computing prowess. And getting to learn about advanced technology even earlier in this stage was an interesting part of this course altogether.''
\end{displayquote}

Positive feedback highlights the effectiveness of structured preparation and interactive elements, such as student presentations and tools like Kahoot. The absence of negative comments on the HL lectures further suggests that this format is engaging and effective, making it a preferred instructional format for enhancing the overall learning experience.

In summary, the statistical and qualitative analysis of multiple-choice and open-ended lecture evaluation responses indicates that both FLL and HL groups received positive feedback across both undergraduate and graduate levels. Overall, in the open-ended responses, students express a strong preference for and engagement in active learning activities.  %

\subsection{Summary of Findings}

Overall, the pilot results suggest that the lightweight PQC module was well received in both undergraduate and graduate contexts. Kahoot-based formative assessment results provide preliminary evidence that the HL format may support short-term conceptual understanding, particularly in the graduate cohort, while the undergraduate results show a positive but non-significant trend. Student feedback further suggests that students found the PQC topic engaging and appreciated the interactive components, including student-led presentations, hands-on activities, and Kahoot quizzes.

These findings support the feasibility of embedding a short PQC module into existing computing courses. However, as discussed in Section~\ref{subsec:data}, because the pilots involved small cohorts and a short one-week module, the results should be interpreted as exploratory evidence of student engagement and short-term conceptual understanding rather than as a controlled evaluation of instructional effectiveness. The primary value of these pilots lies in the design insights and practical lessons they provide for future PQC curriculum development.

\section{Discussion}\label{sec:takeaways}

This section summarizes the main lessons learned from designing, deploying, and iterating the PQC plug-in module. Rather than treating the pilots as a controlled study, we use the results and classroom observations to identify practical design considerations for educators who wish to introduce PQC into existing cybersecurity, software engineering, or computing curricula.

\textbf{Scaffold interdisciplinary PQC concepts for heterogeneous students.}
PQC sits at the intersection of cryptography, mathematics, programming, and quantum concepts, and few students enter the classroom with balanced preparation across all these areas. We observe that students with limited exposure to cryptography or quantum computing often struggle to connect technical definitions to the broader motivation for PQC. Therefore, short PQC modules should make prerequisite knowledge explicit, revisit basic asymmetric cryptography and threat models, and carefully sequence quantum concepts such as qubits, Shor's algorithm, and Grover's algorithm before introducing concrete PQC schemes and migration scenarios. This scaffolding is especially important when the module is embedded into existing courses with diverse student backgrounds.

\textbf{Use short hands-on and gamified activities to make PQC concrete.}
Purely conceptual lectures risk making PQC appear abstract or distant from software practice. In our pilots, short factorization exercises, live demonstrations, and Kahoot quizzes helped students connect quantum threats to familiar cryptographic assumptions and secure software systems. These activities did not need to be large projects; even 10--20 minute guided exercises helped students see why RSA depends on classical hardness assumptions and why PQC migration matters. At the same time, gamified assessment should be used as a complement to structured explanation, not a replacement for it. Kahoot is useful for engagement and immediate feedback, but it measures short-term conceptual understanding rather than deep PQC competency.

\textbf{Use student-led preparation selectively, especially for diverse graduate cohorts.}
The hybrid lecture format asks students to prepare short ``elevator-pitch'' presentations on foundational quantum or PQC topics before the faculty-led lecture. This approach helps distribute cognitive load and gives the instructor a foundation for correcting misconceptions and synthesizing key ideas. In our pilots, the graduate HL cohort shows stronger Kahoot performance than the graduate FLL cohort, while the undergraduate difference is positive but not statistically significant. This suggests that student-led preparation may be especially useful for heterogeneous graduate or professional cohorts, where students benefit from additional self-directed preparation time. However, this approach requires clear topic lists, expectations, and example resources to avoid overloading students.

\textbf{Frame PQC as security engineering and support reuse with adaptable materials.}
Students respond more strongly when PQC is framed not only as a cryptography topic but also as a software evolution and security engineering challenge. Anchoring the module in migration scenarios, legacy systems, and ``harvest-then-decrypt'' risks help students understand why PQC matters for systems they may design, maintain, or audit. Making slides, recordings, and exercises publicly available also lowers the barrier for reuse. However, ``plug-and-play'' adoption is unrealistic: instructors still need to adapt materials to their local curriculum, programming background, course length, and student preparation. Therefore, reusable PQC education resources should document prerequisites, suggested insertion points, and known pain points in addition to providing lecture materials.

\textbf{Limitations.}
This work should be interpreted as an experience report and pilot study rather than a controlled evaluation of PQC learning outcomes. First, the module is intentionally short, lasting approximately one week, and is designed to raise awareness and build foundational understanding rather than develop deep technical mastery of PQC. Second, the study involves small cohorts from a single institution, which limits generalizability. Third, the FLL and HL groups are drawn from different course offerings rather than randomized experimental and control groups, so observed differences may be influenced by cohort composition, semester-level variation, or students' prior preparation. Fourth, the Kahoot quiz is used primarily as a formative assessment and engagement tool; although it provides useful evidence of short-term conceptual understanding, it does not measure long-term retention, implementation ability, or deeper PQC competency. Finally, student feedback is self-reported and may reflect students' immediate perceptions of engagement rather than durable learning gains.

Despite these limitations, the pilots offer practical value for educators and curriculum designers. First, the work demonstrates the feasibility of introducing PQC through a short plug-in module that can be embedded into existing cybersecurity and software engineering courses without requiring a new standalone course. Second, it provides a reusable instructional structure that combines concise lectures, hands-on exercises, student-led preparation, and gamified formative assessment to support engagement with interdisciplinary PQC concepts. Third, the study identifies concrete design considerations, including the need to scaffold prerequisite knowledge, adapt the module to different student backgrounds, and frame PQC as both a cryptographic and software engineering challenge. Finally, by releasing the instructional materials and artifacts, this work lowers the barrier for other instructors to adopt, adapt, and evaluate PQC education in their own contexts. Thus, while the findings should be interpreted as preliminary, the experience reported here provides actionable guidance and a foundation for future, larger-scale curriculum studies.

\section{Conclusions and Future Work}\label{sec:conclusions}
This paper presented an experience report on designing and piloting a one-week lightweight PQC plug-in module for higher-education computing curricula. The module was embedded in an undergraduate cybersecurity course and a graduate software engineering course, allowing us to explore how PQC concepts can be introduced in different curricular contexts without requiring a standalone course. Across four pilot deployments, students generally responded positively to the module, and the results provide preliminary evidence that active-learning strategies, especially student-led seminars combined with faculty-led synthesis, may help students engage with interdisciplinary PQC concepts.

We view this work primarily as a reusable curriculum design and experience report rather than a definitive evaluation of PQC learning effectiveness. The main value of the study lies in its practical design guidance, open instructional materials, and lessons learned for educators seeking to introduce quantum-safe cybersecurity topics into existing computing courses.

Future work will expand the module into a longer online course, incorporate richer hands-on activities and case studies, and evaluate learning outcomes using more comprehensive assessment instruments. These efforts will support the development of quantum-aware software engineers and cybersecurity professionals prepared for post-quantum migration challenges.

\section*{Data Availability}

The dataset, program code, and research materials are publicly available on Zenodo as part of an open-access project (accessible at \url{https://doi.org/10.5281/zenodo.13909016}). In addition, we have published our slides and lecture recordings online at \url{https://est.umbc.edu/teaching/pqc-teaching-module}.

\section*{Acknowledgement}

This work is supported by the Hrabowski Innovation Fund Seed Award at the University of Maryland, Baltimore County. 

\bibliographystyle{IEEEtran}
\bibliography{references}

\end{document}